\pdfoutput=1

\documentclass[reprint,amsmath,amssymb,floatfix,aip,longbibliography]{revtex4-1}
\usepackage{graphics,graphicx,float}

\usepackage[pdftex,
            pdfauthor={Steven A. Frank},
            pdftitle={},
            pdfsubject={Generative models versus underlying symmetries to explain biological pattern}]{hyperref} 

\def\citeyear{\citep}
\def\autocite{\citep}

\newcommand{\Fhat}{\hat{F}}
\newcommand{\mn}{M_n}
\newcommand{\Fr}{Fr\'echet}

\newcommand{\Ga}{\alpha}
\newcommand{\Gb}{\beta}

\newcommand{\Gg}{\gamma}

\newcommand{\dd}{{\hbox{\rm d}}}

\newcommand{\Eq}[1]{eqn~\ref{eq:#1}}

\newcommand{\ovr}[2]{{\frac{#1}{#2}}}
\newcommand{\dovr}[2]{\ovr{\dd #1}{\dd #2}}

\newcount\BoxNum \BoxNum 1\relax
\makeatletter
\newcommand{\boxlabel}[1]{%
  \protected@write \@auxout {}{\string \newlabel {box:#1}{{\the\BoxNum}}{}}%
  \advance\BoxNum 1\relax}
\makeatother

\begin{document}

\title{Generative models versus underlying symmetries to explain biological pattern}

\author{Steven A.\ Frank}
\email[email: ]{safrank@uci.edu}
\homepage[homepage: ]{http://stevefrank.org}
\affiliation{Department of Ecology and Evolutionary Biology, University of California, Irvine, CA 92697--2525  USA}

\begin{abstract}

Mathematical models play an increasingly important role in the interpretation of biological experiments. Studies often present a model that generates the observations, connecting hypothesized process to an observed pattern. Such generative models confirm the plausibility of an explanation and make testable hypotheses for further experiments. However, studies rarely consider the broad family of alternative models that match the same observed pattern. The symmetries that define the broad class of matching models are in fact the only aspects of information truly revealed by observed pattern. Commonly observed patterns derive from simple underlying symmetries. This article illustrates the problem by showing the symmetry associated with the observed rate of increase in fitness in a constant environment. That underlying symmetry reveals how each particular generative model defines a single example within the broad class of matching models. Further progress on the relation between pattern and process requires deeper consideration of the underlying symmetries\footnote{\href{http://dx.doi.org/10.1111/jeb.12388}{doi:\ 10.1111/jeb.12388} in \textit{J. Evolutionary Biology}}$^,$\footnote{Reprint at \href{http://stevefrank.org/reprints-pdf/14JEBsymm.pdf}{http://stevefrank.org/reprints-pdf/14JEBsymm.pdf}}.

\end{abstract}

\maketitle

\section*{Introduction}

In a recent experiment, bacterial fitness increased steadily over time \autocite{wiser13long-term}. The logarithm of the rate of increase in fitness declined approximately linearly with respect to the logarithm of time. Such a simple pattern contains information about the underlying processes that determine how fitness increases. But exactly what sort of information?

To evaluate the match between an observed pattern and a hypothesized process, mathematical models have become the standard in biology. Typically, one puts together a set of plausible assumptions about process, and then studies the resulting model for how well it generates the target outcome. A successful match implies a plausible generative model of process. Ideally, the generative model will make additional testable hypotheses, which can be studied in further experiments. 

But does a successful generative model, by itself, really provide much information about underlying process? Probably not. The more commonly a pattern is observed, the more important it is to understand the underlying process. At the same time, it is almost always true that the more common a pattern, the greater the number of underlying generative models that match the pattern. The simple law of nature is that the commonness of a pattern associates with the number of distinctive underlying processes that lead to that pattern \autocite{jaynes03probability}. Put another way, it is overwhelmingly easy to make a generative model that matches a simple, common pattern, but the match provides little information about the true underlying process \autocite{frank09the-common}.

What is the real information in a pattern about underlying process? The real information is the constraints that define the underlying class of matching generative models \autocite{jaynes03probability}. Such constraints express the fundamental symmetries that determine pattern \autocite{feynman67the-character,anderson72more}. In this case, symmetry means the group of alternative models that produce the same pattern. Symmetry defines alternative states that are the same with regard to some measure \autocite{weyl83symmetry}. For example, a square is the same after it is rotated by ninety degrees, so a square is symmetric with respect to ninety degree rotations. Similarly, one may change many assumptions of an evolutionary model and still generate a linear decline in the logarithm of fitness increase with respect to the logarithm of time. The rate of increase in fitness is symmetric with respect to such changes in underlying generative assumptions. 

If we knew all of the symmetries with respect to the rate of increase in fitness, then we would know the full class of underlying generative models consistent with that pattern. We would know the essence of process required to generate the pattern, and those aspects of particular generative models that do not matter. For simple patterns, the associated symmetries tend to be simple, and most details of particular generative models do not matter.  However, we can only distinguish those aspects that matter from those that do not if we know the defining symmetries.

Why are most studies limited to expressing the existence of a matching generative model? Because finding a matching model is easy, whereas discovering the underlying symmetries that truly define the relations between pattern and process is often difficult. In consequence, it has become common to ignore the inherent structure of the problem, and to be satisfied with a generative match. Indeed, the very idea that one should look for underlying symmetries rather than a generative match is rarely acknowledged and perhaps not widely understood.

I do not have a solution to the difficulty of discovering the underlying symmetries. However, to have a chance, one must first recognize the problem. With the proper goal in mind, certain steps often help in learning about the underlying symmetries of process that lead to observed pattern. In this article, I use the example of increasing bacterial fitness to illustrate some of these issues.

\section*{Overview}

The argument in this article is a bit more abstract than usual. A brief overview may help before starting. A common sequence of science is hypothesis, test by observation, and updated hypothesis in light of observed pattern. How well one does with that sequence depends on how good one is at finding useful updated hypotheses. An updated hypothesis must, of course, be constrained by the observed pattern. But consistency with observed pattern may often be, by itself, a rather weak way of generating new hypotheses. 

For example, if one observed a Gaussian (Normal) distribution of measurements, then a detailed hypothesis about why the specific natural history or biochemistry of certain processes led to that Gaussian pattern could easily be constructed to generate the pattern. But such a detailed explanation of process would almost certainly be wrong, because the central limit theorem tells us that Gaussian patterns arise inevitably when underlying processes tend to add up to make final pattern, irrespective of almost all of the details concerning the underlying processes. 

The phrase ``irrespective of almost all of the details concerning the underlying processes'' means that pattern in that case is symmetric or invariant to many changes in details. One must know that, otherwise the hypothetico-deductive process is almost certain to lead one to false paths, because one may mistakenly put too much weight on hypothesized details of process that in fact do not matter. The same problem arises for many patterns that are not Gaussian in shape. The difficulty is that the underlying symmetries are not always immediately obvious, and so require some thought in order to avoid false paths. To illustrate those points, I have structured this article as follows.

The first section introduces a particular observed pattern for the increase in fitness over time in bacterial populations. I express that observed pattern in a variety of alternative ways. Those alternative expressions allow one to see the pattern from a variety of perspectives. Those different perspectives provide a deeper sense of the pattern and its ``shape.''  A sense of shape helps to see what may matter and may not with regard to underlying process, that is, the underlying symmetries.

The second and third sections introduce the particular symmetries that may explain the form of the observed pattern for the increase in fitness. In this case, the symmetries arise from extreme value theory. That theory has the same structure as the central limit theorem. In particular, a wide variety of seemingly different processes turn out to lead to the same pattern, when the pattern depends primarily on rare or extreme events. At first glance, it may seem that rare events would be particularly difficult to predict, and so be difficult to analyze with regard to the consequences for pattern. However, although each rare event is hard to predict, in the aggregate over several rare events, the outcomes tend to converge to a very regular form. One must recognize that symmetry in order to find meaningful hypotheses about the generation of certain types of pattern.

The fourth section applies extreme value theory to the observed pattern for the increase in fitness in bacterial populations. The final sections interpret the particular results with respect to the broader problem of understanding symmetries and their role in generating useful hypotheses for the interpretation of commonly observed patterns.

\section*{Observed power law scaling for the increase in fitness}

\textcite{wiser13long-term} studied the change in fitness over time in experimental populations of bacteria maintained in a constant environment. They showed that the observed fitnesses over time follow a power law relation
\begin{equation}\label{eq:origPower}
 w = (1+bt)^a,
\end{equation}
where $w$ is mean fitness at time $t$, estimated from 12 replicate populations, and $a$ and $b$ are fitted parameters. For my purposes, it will be useful to express Wiser et al.'s relation in a variety of alternative ways to highlight the related forms of simplicity that define the essential pattern. Most of the simple forms begin with logarithmic scaling
\begin{equation}\label{eq:fitPower}
 \log w = a\log(1+bt),
\end{equation}
which, for $bt \gg 1$, approaches
\begin{equation}\label{eq:fitPower2}
 \log w = a\log b + a\log t.
\end{equation}

\subsection*{Rate of change in mean fitness}

From \Eq{fitPower2}
\begin{equation}\label{eq:dlogPower}
 \dovr{\log w}{\log t} = a,
\end{equation}
which shows a very simple relation between the change in log mean fitness and the change in the logarithm of time.

We can also write the change in log fitness with respect to time as
\begin{equation}\label{eq:dlogPower2}
 \dovr{\log w}{t} = \Gg
\end{equation}
where $\Gg\equiv\Gg(t)$ is a function of time that describes the rate of increase in log mean fitness at any time. Note that if selection is the only force, then $\Gg\equiv J$ from \textcite{frank12naturalc}, and $\Gg$ is the increase in information accumulated by a population from the action of natural selection. 

It follows from \Eq{fitPower} that the increase in information at any point in time is
\begin{equation}\label{eq:gamma}
 \Gg = \frac{ab}{1+bt} \rightarrow \frac{a}{t},
\end{equation}
where the limiting value on the right arises when $bt\gg 1$.

\subsection*{Decay in the rate of change in log fitness}

From \Eq{gamma}, it follows that
\begin{equation}\label{eq:gammaScaling}
 \dovr{\log \Gg}{\log t} = -1,
\end{equation}
which shows that the logarithm of the rate of increase in fitness declines directly in proportion to $\log t$, the logarithm of the amount of time that has passed.

\subsection*{Time required for a fixed change in log fitness}

We can write \Eq{dlogPower} as
\begin{equation}\label{eq:dloglogPower}
 \dovr{\log w}{\log t} = \Gg t = a,
\end{equation}
because $\dd t = t\dd\log t$, and from \Eq{gamma}, $\Gg t = a$.

How much time must pass for a fixed change in log fitness, $\dd\log w = k$? From \Eq{dloglogPower},  
\begin{equation}\label{eq:logtimeneeded}
 \dd\log t = k/a.
\end{equation}
Because $\dd\log t = \dd t/t$, we have
\begin{equation}\label{eq:timeneeded}
 \dd t = kt/a \propto t,
\end{equation}
where the symbol $\propto$ means "is proportional to." Thus, the time increment, $\dd t$, needed to obtain a fixed change in log fitness is proportional to the amount of time, $t$, that has passed since the first measurement of fitness.

\subsection*{Interpretation of simple scaling relations}

Observation of such simple and elegant scaling relations implies a powerful underlying force. That underlying force must erase all the details of selection and evolution that are particular to each population, exposing the minimal symmetry that constrains pattern. No particular generative or dynamical model can make a primary claim to explaining such simplicity. Rather, one must search for the way in which aggregation and the loss of the particular information in each population causes the simple underlying symmetry to dominate \autocite{frank09the-common}. 

\section*{Extreme values and the increase in fitness}

Several prior studies have analyzed the accumulation of beneficial mutations in a constant environment. Fitness increases over time as each beneficial mutation gets added to a population by selection. The most interesting studies with regard to underlying symmetries emphasize extreme value theory \autocite{gillespie83a-simple,gillespie84molecular,gillespie94the-causes,orr03the-distribution,orr10the-population,beisel07testing,joyce08a-general}. That theory describes the probability distribution for the maximum observed value in a sample. 

For the case of increasing fitness, one can think of a sample as the mutations that occur during a fixed time period. The most beneficial mutation that spreads through a population during a particular period would often be the extreme value, which would determine the advance in fitness over that time increment. The overall rate of increase in fitness depends primarily on the sequence of extreme values over sequential time periods.

The prior studies used extreme value theory to show that many details of underlying models do not matter with regard to the distribution of the effects of beneficial mutations. Because the distribution of beneficial mutations influences the rate of increase in fitness, many details of underlying models do not matter with regard to the increase in fitness. Here, I build on the insight of those prior applications of extreme value theory. In particular, I use extreme value theory to expose the underlying symmetries that may explain why the observed pattern of increase in fitness follows the simple scaling law observed by \textcite{wiser13long-term}. 

\section*{Extreme value theory}

This section summarizes extreme value theory. I limit my presentation here to essential aspects that illustrate the concepts. \textcite{frank09the-common} provides a full introduction and citations to comprehensive treatises on extreme value theory.

I begin with a few definitions. Write the probability that a random process, $Y$, takes on a value above the given constant $y$, as 
\begin{equation*}
  P(Y>y) = \Fhat(y).
\end{equation*}
This expression describes the probability of observing a value in the upper tail of the distribution, that is, a value greater than $y$. However, our primary interest concerns the maximum value, $\mn$, in a sample of size $n$, which we may also think of as the extreme value of a sample. The probability that $\mn$ is less than $y$ is the probability that none of the $n$ samples is greater than $y$, which is
\begin{equation}\label{eq:probMn}
  P(\mn<y) = \left[1-\Fhat(y)\right]^n.
\end{equation}
This expression is the right idea, because it tells us about the probability that the extreme value in a sample is bounded by a particular upper value, $y$. However, there is a mathematical problem. As the sample size, $n$, becomes large, the probability that the maximum value is bounded by $y$ becomes zero, as long as there is some chance that an observation could be above $y$. In a big enough sample, eventually some observation will be above the bound $y$. So this expression by itself is not very helpful with regard to general principles, because the expression depends on the sample size, $n$, which will vary in each particular study.  

To obtain a useful general expression for extreme values, we need to find a form that does not depend on sample size. Put another way, we want an expression for the probability of observing a particular extreme value that is symmetric, or invariant, with respect to changes in sample size. If we can find that underlying symmetry, we can greatly expand our understanding of the general underlying principles that determine the distribution of extreme values. 

The form of \Eq{probMn} suggests that the following mathematical identity will help in finding an expression that is independent of sample size
\begin{equation}\label{eq:limitingform}
  \left[1-\frac{\Fhat(y)}{n}\right]^n \longrightarrow e^{-\Fhat(y)},
\end{equation}
in which the right hand side does not depend on the sample size, $n$, and is an increasingly good approximation for the left hand side as $\Fhat(y)/n$ decreases. The problem now becomes how to express the probability for the maximum in a sample, given in \Eq{probMn}, in the form given by \Eq{limitingform}.  In particular, the problem is that the maximum of sample, $\mn$, depends on the sample size, $n$. How can we standardize $\mn$ to remove the dependence on sample size?  One possibility is to account for the increase in the expected maximum value with $n$ by using the value for the upper bound $\Ga_n y + \Gb_n$, in which the coefficients $\Ga_n$ and $\Gb_n$ depend on $n$ in a way that accounts for the expected increase in the upper bound. In particular, if we can find values of $\Ga_n$ and $\Gb_n$ such that 
\begin{equation*}
  \Fhat(\Ga_n y + \Gb_n) = \frac{\Fhat(y)}{n},
\end{equation*}
then we can rewrite the form in \Eq{probMn} as
\begin{equation}\label{eq:probMnS}
  P(\mn<\Ga_n y + \Gb_n) = \left[1-\frac{\Fhat(y)}{n}\right]^n \longrightarrow e^{-\Fhat(y)}.
\end{equation}

The probability of obtaining a particular extreme value depends on how $\Fhat(y)$, the upper tail probability of observing a value above $y$, declines with an increase in $y$. One commonly observed pattern is a power law decline in the upper tail probability with increasing $y$, such that the expected probability in the upper tail is approximately proportional to $\Fhat(y) = y^{-1/a}$. 

Given a power law scaling in the upper tail, we must choose $\Ga_n=n^a$ and $\Gb_n= 0$ in order to satisfy \Eq{probMnS}, because $\Fhat(n^a y)=\Fhat(y)/n$. Those constants lead to the extreme value distribution of the \Fr\ form
\begin{equation*}
  P(\mn< n^a y) =  e^{-y^{-1/a}}.
\end{equation*}

For our purposes, we only need to use the fact that, for a sufficiently small $\Fhat(y)/n$, the expected value of the maximum, $\mn$, increases with the sample size $n$ in proportion to $n^a$. Using angle brackets to denote expected values, we can express this key fact in symbols as
\begin{equation}\label{eq:expectedlog}
  \langle\mn\rangle = n^a\langle y\rangle,
\end{equation}
under the assumption that $a<1$, so that $\mn$ increases at a diminishing rate with $n$.

\section*{The increase in fitness}

Our goal is understand the observed power law scaling between relative fitness and time. That scaling was given in \Eq{origPower}, repeated here for convenience
\begin{equation}\label{eq:fitPower2repeat}
 w = (1+bt)^a.
\end{equation}
My claim is that \Eq{expectedlog} captures the essence of the observed power law scaling for fitness. To support that claim, we must fill in some gaps to show how the extreme value result in \Eq{expectedlog} expresses the simple underlying symmetries that lead to \Eq{fitPower2repeat}. 

To start, we must show the essential equivalence of the extreme value theory expression in \Eq{expectedlog} and the observed scaling of fitness with time in \Eq{fitPower2repeat}. Then we must express the underlying symmetries in the extreme value theory, and how those symmetries may clarify the types of generative processes that could lead to the observed pattern.

To establish the equivalence, we start by letting $\langle\mn\rangle\equiv w$, where $w$ in this context is the expected relative fitness. The idea is that the best potentially selected mutant, $\mn$, has spread through the population and is the maximum observation, or extreme value. Other mutants may have had a higher value but, for whatever reason, were constrained from spreading and do not enter into the set of potentially selected mutants. 

The expression $\langle\mn\rangle$ is the expected maximum value as a function of the sample size, $n$. We will interpret the sample size as an increasing function of the time that has passed. The relative fitness, $w$, at any time, is equivalent to the best mutant that has spread through the population up to that time.  

The equivalence between fitness and the extreme value means that 
\begin{equation*}
 w = \langle\mn\rangle = n^a\langle y\rangle.
\end{equation*}
Because $w$ is relative fitness, which we may scale by any constant, we may choose $\langle y\rangle\equiv 1$ without loss of generality. We thus obtain the expression for the expected value of relative fitness
\begin{equation}\label{eq:wn}
 w = n^a.
\end{equation}
This very simple result requires only that the probability distribution for potentially selected mutants has an upper tail that decays as a power law. A power law decay in the upper tail for fitness, $w$, is equivalent to an exponentially decaying upper tail when values for the mutants are expressed on a logarithmic scale of fitness, $\log w$. The actual shape of an upper tail is a difficult empirical problem \autocite{clauset09power-law}. A power law may be a good approximation, because fitness has a natural tendency to scale logarithmically \autocite{wagner10the-measurement}. Logarithmic scaling typically corresponds to power laws \autocite{frank09the-common}. For specific discussion of alternative tail shapes of mutational effects in relation to extreme value theory, see the excellent analysis in \textcite{joyce08a-general}.

The result in \Eq{wn} describes fitness in terms of sample size, $n$. However, the observed pattern of fitness in \Eq{fitPower2repeat} expresses the change of fitness with respect to time, $t$, rather than with respect to sample size, $n$. Thus, we must define a reasonable relation between sample size, $n$, and time, $t$. The relation that transforms \Eq{wn} into \Eq{fitPower2repeat} is
\begin{equation}\label{eq:nt}
 n=1+bt.
\end{equation}
This equivalence works mathematically, but can we justify it biologically? In fact, the equivalence is inevitable if we take the simplest interpretation of two necessary relations. First, at $t=0$, the biologically observed expression in \Eq{fitPower2repeat} uses the arbitrary assumption that relative fitness is one. Using that same assumption, we must have $t=0$ corresponding to $n=1$, for which \Eq{nt} uses the simplest expression of that assumption. Second, if we follow the inevitable fact that sample size increases with time, then the simplest assumption is that the number of samples increases linearly with time. In \Eq{nt}, the assumption is that the sample size, $n$, increases linearly with time as $bt$. For these two reasons, \Eq{nt} seems the simplest, fully justified way to express the relation between sample size and time. 

Using the relation between sample size and time from \Eq{nt} in the universal extreme value scaling law in \Eq{wn}, we obtain the observed pattern for the change of fitness with time in \Eq{fitPower2repeat}. The only assumptions are: upper tail events are relatively rare; upper tail events that increase fitness decay as a power law; and sample size increases linearly with time. The observed power law pattern contains only the information in those three assumptions. For any generative process that matches the observed pattern, no additional information can be added beyond those key assumptions. 

\section*{Symmetry interpretation versus generative models}

For the observed power law increase of fitness with time in \textcite{wiser13long-term}, all matching generative models are symmetric with regard to details beyond the two key aspects of information contained in the observed pattern. To repeat the above conclusion, the key aspects of information are: upper tail events that increase fitness decay as a power law; and upper tail events are relatively rare \autocite{gillespie94the-causes,joyce08a-general,orr10the-population}. Here, symmetry means invariance, in the sense that the pattern generated by a matching model is invariant with regard to any details beyond the two components of information contained in the observed pattern.

The central limit theorem provides the best known example of symmetry and invariance \autocite{jaynes03probability}. When summing up a series of random processes, one often observes a normal, or Gaussian, distribution. The reason is that the particular details in each component process tend to average out, exposing only the underlying information about the mean and variance of the aggregate process. Thus, the mean and variance are sufficient statistics to define a Gaussian distribution. Put another way, given the mean and the variance, all other details of a particular generative process get washed away in the aggregate. The Gaussian pattern is symmetric, or invariant, to the other particular details of the great many generative processes that lead to that outcome. Indeed, the Gaussian is so very common because there are so many different generative processes that satisfy the key underlying symmetry.

The central limit theorem is well known. Thus, an observed Gaussian pattern rarely motivates anyone to produce a detailed generative model to match that pattern. The fact that so many different generative models lead to that pattern is understood. By contrast, in biology, almost any other observed pattern often motivates the production of a detailed generative model. But many other patterns besides the Gaussian share the essential feature of attracting a wide class of underlying generative processes to the same pattern. The extreme value pattern is one of the most powerful attractors, on the same level as the Gaussian pattern \autocite{embrechts97modeling,kotz00extreme,frank09the-common,frank11a-simple}. If the set of attracting patterns were better known, the widespread tendency of explaining simple general patterns by overly particular and often misleading generative models would not be so common \autocite{gutenkunst07universally,frank11measurement,frank13input-output,machta13parameter}. 

The philosophical literature contains a distinct analysis of similar problems. As \textcite{stanford13underdetermination} puts it: ``At the heart of the underdetermination of scientific theory by evidence is the simple idea that the evidence available to us at a given time may be insufficient to determine what beliefs we should hold in response to it.'' Focus on symmetries provides a way forward, by telling us something about the particular nature of underdetermination.

\section*{Discussion}

In this section, I first provide additional context about symmetries in relation to explanation. I then discuss how my approach to the particular example of fitness relates to past work.

Underlying symmetries do not explain all of the details of observed pattern. Rather, the symmetries express the main force that sets the broad features of pattern. In the same way, the average value over a range of outcomes does not express all of the variability, but rather the expected outcome or the general location of pattern. Thus, my argument is not that the details of particular models are necessarily uninteresting or unhelpful. Instead, one must first locate the primary cause of pattern before one can understand the causes of variation around that primary location. The first step in explanation is to figure what does not matter in setting the main shape of pattern. One may then fill in the details of how, in particular situations, additional processes set the variations around the primary shape.

Often, a complicated and detailed generative model, which appears to capture many aspects of realism, attracts strongly to the very simple pattern set by the basic underlying symmetries. The many details act like random perturbations relative to the core pattern. The greater the number of random perturbations, the more likely they tend to cancel in the aggregate, leaving only the core pattern.

\textcite{wiser13long-term} present a detailed generative model to explain their observed power law scaling of fitness with time. They assume that beneficial mutations follow an exponential distribution when measured with respect to $\log w$, which is equivalent to a distribution with a power law upper tail for $w$. They also make many particular assumptions, leading to a relatively complicated model and analysis. Their primary conclusion is that their model matches the simple power law pattern for the increase of fitness with respect to time. They also include analysis and discussion of many potentially realistic and informative details of evolutionary process. Those details could provide much insight. The main limitation is that, by not expressing the underlying symmetry and the simplicity of the relation between pattern and process, it is hard to see what matters and what does not matter with respect to the primary pattern. Therefore, it is hard to know how the particular details do or do not influence variation around the primary trend set by the underlying symmetry.

Much past work has promoted the power of extreme value theory for analyzing the distribution of beneficial mutations and the rate of increase in fitness \autocite{gillespie94the-causes,joyce08a-general,orr10the-population}. That work does emphasize the importance of underlying symmetries in understanding the relation between pattern and process. However, the particular analyses from the past work have sometimes focused on rather detailed and specific patterns or assumptions \autocite{joyce08a-general}. Those details could, in theory, matter a lot when analyzing the actual patterns of mutation, selection, and the increase in fitness. 

The great value of the data and analysis from \textcite{wiser13long-term} is the simplicity of observed pattern. My main goal has been to match that observed simplicity to a theory that captures the underlying symmetry, in a way that could explain the primary cause of such simplicity of pattern. Of course, other general models besides extreme value theory could match the same observed pattern. But that is exactly my point. Simple, common patterns tend to be attractors for many underlying models. Extreme value theory does have a privileged position, shared only with the central limit theorem and its generalization to a broad family of related patterns \autocite{frank09the-common,frank11measurement,frank11a-simple}. Many common patterns arise from those few families of special attractors. 

\section*{Acknowledgments}

National Science Foundation (USA) grants EF--0822399 and DEB--1251035 support my research.  

\bibliography{main}

%
%
%
%


\end{document}